# Tuning of the Dzyaloshinskii-Moriya Interaction by He[+] ion irradiation

Hans T. Nembach[1,2,*], Emilie Jué[2,3], Kay Poetzger[4], Juergen Fassbender[4], Thomas J. Silva[2], Justin M. Shaw[2]

[1]JILA, University of Colorado, Boulder, Colorado 80309, USA

[2]Quantum Electromagnetics Division, National Institute of Standards and Technology, Boulder, Colorado 80305, USA

[3]Department of Physics, University of Colorado, Boulder, Colorado 80309, USA

[4]Institute of Ion Beam Physics and Materials Research, Helmholtz-Zentrum Dresden - Rossendorf, Dresden, Germany

[*]hans.nembach@nist.gov

We studied the impact of He[+] irradiation on the Dzyaloshinskii-Moriya interaction (DMI) in Ta/Co$_{20}$Fe$_{60}$B$_{20}$/Pt/MgO samples. We found that irradiation 40 keV He[+] ions increases the DMI by approximately 20 % for fluences up to $2 \times 10^{16}$ ions/cm$^2$ before it decreases for higher fluence values. In contrast, the interfacial anisotropy shows a distinctly different fluence dependence. To better understand the impact of the ion irradiation on the Ta and Pt interfaces with the Co$_{20}$Fe$_{60}$B$_{20}$ layer, we carried out Monte-Carlo simulations, which showed an expected increase of disorder at the interfaces. A moderate increase in disorder increases the total number of triplets for the three-site exchange mechanism and consequently increases the DMI. Our results demonstrate the significance of disorder for the total DMI.

*Introduction* - The Dzyaloshinkii-Moriya interaction (DMI) is an anti-symmetric exchange interaction that requires broken inversion symmetry [1–3]. The DMI can originate from interfaces[4,5], where a ferromagnet is, for example, in contact with a heavy metal like Pt or Ir , or from the bulk itself, such as for the weak ferromagnet FeBO$_3$, or for hematite, Fe$_2$O$_3$ [4]. Recently, it has been shown that DMI is also present at the interface between a ferromagnet and graphene [5] or an oxide [6]. DMI gives rise to a canted alignment of adjacent spins and chiral spin-structures, e.g., skyrmions. The latter are protected by their topological spin texture, making skyrmions viable candidates for nonvolatile magnetic memory technologies [7–9]. DMI can also influence the switching behavior in more conventional magnetic random-access memory devices, where it can reduce thermal stability and increase switching currents [10,11].

Not much is understood about the influence of interface morphology on DMI. So far, the importance of spin-orbit coupling and symmetry-breaking has been primarily emphasized. However, it has already been shown that a nominally symmetric structure, where a ferromagnetic layer is sandwiched between two Pt layers, can still exhibit a net DMI, even though the DMI for the two interfaces should be equal and opposite [12–14]. Wells *et al.*, studied the effect of substrate temperature and chamber base pressure during deposition of Pt/Co/Pt thin films on DMI. They showed that the DMI is largest when the interface quality for the top and bottom interfaces differs the most [12].

Ion irradiation is a powerful tool for altering magnetic properties [15–17]. Balk *et al.*, demonstrated that the sign and magnitude of the resulting DMI in such structures can be manipulated, and even reversed, by irradiation with Ar$^+$ ions [18]. This implies that both interfaces have been altered differently by the ions such that the relative magnitude of DMI changed. Recent work by L. Diez *et al.* reported on the irradiation of Ta/CoFeB/MgO samples with He$^+$ ions. In that work, domain wall velocity and Brillouin light scattering (BLS) measurements suggested the total DMI was enhanced by ion irradiation [19]. However, the possibility that this was due to a reduction in the thickness of the magnetic layer as a result of the irradiation was not ruled out. Further, their BLS data is not consistent with basic models that require a zero intercept for the wavevector dependence of the DMI-induced frequency shift [20,21]. The impact of irradiation with heavy Ga$^+$ and He$^+$ ions on the DMI was reported by R.Gieniusz et al. [22] and J. Jun et al. [23], respectively. In both studies, the DMI does not increase above its initial magnitude with ion-irradiation.

In the present work, we use light ion-irradiation to systematically explore how DMI and the perpendicular anisotropy can be tuned by controlled modification of the ferromagnet/heavy metal interface morphology. We quantify the magnetic anisotropy and a thickness-independent measure of the DMI to isolate the origin of the DMI enhancement. Our results show that the DMI increases by up to 20 % before it decreases for the highest fluence. In contrast, the anisotropy shows a distinctly different trend with ion irradiation.

*Experiment* - We sputter-deposited a Ta(3 nm)/Co$_{20}$Fe$_{60}$B$_{20}$(1.9 nm)/Pt(3.5 nm)/MgO(1 nm) film on a thermally oxidized silicon wafer at room temperature. The deposition rates were calibrated by X-ray reflectometry. Here, the MgO layer is the capping layer, and the Ta is the adhesion layer. The samples were not annealed post deposition. Following deposition, the wafer was diced into chips for ion-irradiation. Since all the samples came from chips diced from the same wafer, the sample-to-sample variation was minimized. Individual chips were irradiated with He$^+$ ions at a fixed energy of 40 keV and a variable fluence of up to $4 \times 10^{16}$ ions/cm$^2$. The ion flux was limited to no more than 500 nA/cm$^2$ to avoid annealing. At this energy, it is expected that all ions fully penetrate through the multilayer film and eventually terminate in the Si substrate [17]. We characterized the samples after ion-irradiation with superconducting quantum interference device (SQUID) magnetometry, ferromagnetic resonance (FMR), and Brillouin light scattering (BLS). In addition, we carried out Monte-Carlo computations to study the impact of the ion-irradiation on the Ta and the Pt respective interface with CoFeB, and to get a better understanding of how interface modifications change the DMI.

The saturation areal magnetic moment, obtained by dividing the total magnetic moment by the sample area, decreases with increasing ion irradiation (Fig. 1). The reduction of magnetization could be due to a decrease of the room-temperature saturation magnetization $M_s$, and/or a reduction in the thickness of the CoFeB layer due to intermixing. A reduction of the anisotropy and the Curie temperature as a result of light-ion irradiation has been previously reported for the Pt/Co/Pt system [17,24]. The latter implies a reduction of $M_s$ at a given temperature.

We determined the perpendicular anisotropy field $H_k$ from FMR measurements in the perpendicular geometry. See Nembach *et al.* for details [25]. We fit the frequency dependence of the resonance field with the Kittel equation $f = \frac{\mu_0 g \mu_B}{h}(H - M_{eff})$, where $\mu_0$ is the permeability of free space, $\mu_B$ the Bohr magneton, $g$ the spectroscopic splitting factor, $M_{eff}$ the effective magnetization and $h$ the Planck constant. The ion-irradiation can reduce the saturation magnetization of the ferromagnet through intermixing with atoms from the adjacent layers and can also reduce the thickness of the ferromagnet. We cannot determine the relative importance of these two changes and have to consider two limiting assumptions to determine the anisotropy from the data. In the first case, we assumed the nominal CoFeB thickness to calculate the magnetization $M_s$ from the magnetic moment obtained from SQUID magnetometry. This combined with the values for M$_{eff}$ are used to determine the anisotropy field for constant thickness $H_k^t = M_s - M_{eff}$. Fig. S1a in the supplemental information shows that the anisotropy field decreases rapidly at first with increasing ion irradiation before the decrease becomes more moderate at a fluence of about $10^{16}$ ions/cm².

In the second case, we instead assumed constant $M_s$ under a scenario where the actual CoFeB thickness *t* is now a decreasing function of fluence. By this assumption, a change in the magnetic moment would indicate a change in the magnetic thickness, due to, for example, the presence of a magnetic dead layer. Here, the anisotropy field $H_k^{Ms}$ initially decreases before it increases for the two highest fluences (see Fig. S1b in the supplemental information). Both assumptions yield a reduction of $H_k$ for most fluences albeit by different amounts. Only for the two highest fluences the two assumptions lead to opposite trends for $H_k$. Consequently, a detailed knowledge of the fluence dependence of *t* and $M_s$ is required to make any conclusions about $H_k$.

We calculated the DMI from the non-reciprocal frequency-shift $\Delta f$ of Damon-Eshbach spinwaves measured with BLS. These spinwaves propagate in the film-plane and perpendicular to the direction of the magnetization. When the direction of the magnetization or the propagation direction of the spinwaves is reversed, the sign of $\Delta f$ changes[21–26]. We determined $\Delta f$ by measuring the frequency of the Stokes and the anti-Stokes peaks for both field polarities. See Fig. 2a for the spectra of a sample before irradiation. The DMI for this material system is an interfacial effect. As such, the physically relevant quantity that is independent of film thickness *t* is the interfacial $D_{int}$, which is related to the volume-averaged DMI value $D = D_{int}/t$. We can calculate $D_{int}$ without any assumptions concerning modifications of the sample thickness and $M_s$ during ion irradiation from the measured $\Delta f$:

$$|D_{int}| = \frac{|\Delta f| h}{2 g \mu_B k} \cdot m_A,$$

where *k* is the spinwave wavenumber and $m_A$ is the measured magnetic moment per unit area of the sample, as shown in Fig. 1a. The $\lambda = 532$ nm laser beam for the BLS measurements was incident at 45 degrees to the sample surface and the detected spinwave wavenumber is $k = 16.7 \ \mu m^{-1} = 4\pi \cos(\pi/4)/\lambda$. We use the sign convention that positive $D_{int}$ promotes clockwise rotation chiral spin structures.

Damon-Eshbach spinwaves have a non-reciprocal amplitude profile throughout the thickness of the ferromagnet. As such, spinwaves propagating in opposite directions are affected by the interface anisotropies of the top and bottom interface differently. This results in a non-reciprocal frequency shift $\Delta f_{ani}$ of the spinwaves. The change in $\Delta f_{ani}$ that is simply due to the radiation-induced change in interfacial anisotropy, independent of any change in the DMI, can be accounted for by use of the following relation from Gladii *et al.* [30]:

$$\Delta f_{ani} \simeq \frac{8\gamma}{\pi^3} \frac{K_1 - K_2}{M_s} \times \frac{k}{1 + \frac{\Lambda^2 \pi^2}{t^2}},$$

where $K_1$ and $K_2$ are the interface anisotropies of the bottom and top interface, respectively, $\Lambda = \sqrt{2A/\mu_0 M_s^2}$ is the exchange length, and $A$ is the exchange energy density. We use the value $A$ = 9.5 pJ/m reported by M. Belmeguenai, *et al.*, for the exchange constant [31] , and we use the nominal thickness $t = 1.9$ nm, which was also used to determine $M_s$. We then calculate the upper bound for the expected anisotropy-driven frequency shift by use of the maximum measured change in the anisotropy from FMR for a constant thickness of the ferromagnet. As both the Pt and the Ta interface can in principle contribute to the perpendicular anisotropy, we then calculate for the case where all the anisotropy originates at a single interface, e.g., $K_1 > 0$ and $K_2 = 0$ in order to obtain an upper bound for $\Delta f_{ani}$. The largest $\Delta f_{ani}$ we calculate is ~2 MHz, which is about 2 % of the maximum change in the frequency-shift of ~100 MHz we measure. We use the calculated upper bound to $\Delta f_{ani}$ in our determination of the measurement uncertainty for the DMI.

*Results:* The results for $D_{int}$ are shown in Fig. 2b. The error for $D_{int}$ is calculated using standard error propagation for the uncertainty in the fits of the respective spectra, the uncertainty in $\Delta f$ due to the interface anisotropy and the errors in *g* and $m_A$. While the scatter in the data is substantial, it is possible to discern that $D_{int}$ has a broad maximum at $2 \times 10^{16}$ ions/cm$^2$, where it is approximately 20% higher than at zero irradiation. For the highest irradiation of $4 \times 10^{16}$ ions/cm$^2$, the DMI drops back toward its original value. The reduction of the DMI for the highest fluence is to be expected as ultimately all symmetry-breaking will disappear and as a consequence the DMI.

Both the Ta as well as the Pt interfaces potentially contribute to the total DMI, though we believe that the Ta interface gives rise to a much smaller DMI than the Pt interface; Chaurasiya *et al.* measured $|D_{int}|$~0.2 pJ/m for Ta/CoFeB/TaO$_x$, where the oxide interface might also contribute to the total DMI [32,33] and Arora *et al.*, found $|D_{int}| = 0.032 \pm 0.015$ pJ/m for Cu/Co$_{90}$Fe$_{10}$/Ta, and $|D_{int}| = 0.049 \pm 0.014$ pJ/m for Cu/Co$_{90}$Fe$_{10}$/TaO$_x$ [34]. We measured $|D_{int}| = 1.11 \pm 0.02$ pJ/m for Ta/CoFeB/Pt without any ion irradiation. The net DMI increase of 0.2 pJ/m for $2 \times 10^{16}$ ions/cm$^2$ is 7 times larger than the DMI for Cu/CoFe/Ta, and the bottom Ta layer in our multilayer induces the opposite chirality of the top Pt layer. These prior results strongly suggest that the majority of the DMI increase is the result of radiation-induced modification of the Pt/CoFeB interface.

We carried out Monte-Carlo simulations for each irradiation fluence to better understand the effects of ion-irradiation on the interface morphology. The dynamic simulations of ion trajectories in the multilayers were performed using the TRIDYN [35]-code [36], which calculates the history of the incident ions and the development of collision cascades as a sequence of binary collisions. The quantification of atomic displacements and thereby the interface mixing depends on the choice of the threshold energy of recoil atom relocation. For this, a universal value of 8 eV is assumed, which has been successful in numerous ion mixing, sputtering, and thin film deposition studies using TRIDYN [37]. The accounting for accumulated effects is particularly important when a large fraction of atoms near the interface are displaced across the nominal boundary between the two materials. Additional details about the TRIDYN simulations are given in the SI. The thickness resolution for the simulations was 0.2 nm, which required us to model the thicknesses of the Pt layer as 3.6 nm and the thickness of the CoFeB layer as 2.0 nm.  In Fig. 3a, atomically sharp composition profiles are shown for Ta, Pt, Co and Fe before irradiation (solid line), then simulated profiles are shown for fluences of 7×10$^{15}$ ions/cm$^2$ (dashed trace) and 4×10$^{16}$ ions/cm$^2$ (dotted trace). The simulations included both the Si substrate and the MgO cap layer, though the effect on those interfaces is not shown for simplicity.

We calculated the relative number of displaced atoms of CoFeB within the first 0.4 nm of Ta and Pt proximate to the original surface of the Ta and Pt layers. The relative fraction of displaced atoms within the 0.4 nm layer is given by $(n_{Co} + n_{Fe} + n_B)/n_{tot}$, where $n_{Co}$, $n_{Fe}$ and $n_B$ are the number of Co, Fe and B atoms, and $n_{tot}$ is the total number of atoms. A much larger number of atoms originating from the ferromagnet are embedded inside the Pt layer than inside the Ta layer (Fig. 3b), with the total percentage of dissimilar atoms in the 0.4 nm thick region in Pt exceeding 50 % for the highest fluence. For a fluence of $1 \cdot 10^{16}$ ions/cm², the number of atoms embedded in the 0.4 nm thick region of the Pt layer is about twice as large as that in the Ta layer. The numbers of Pt and Ta atoms displaced into the first 0.4 nm of CoFeB proximate to the original surface are comparable for a fluence up to $1 \times 10^{16}$ ions/cm², but the number of Ta atoms exceeds that of the Pt atoms for larger fluences.

Based on the concentration profile for the different atomic species through the thickness of the sample, we carried out an additional series of Monte Carlo simulations for a 2-dimensional vertical slice through the sample stack to determine the number of FM-Pt-FM anisotropic triplets, where FM stands for either a Co or Fe atom, see Fig. 4. These triplets are the fundamental building blocks for interfacial DMI [7,38]. For the Monte Carlo simulations, each atomic layer was represented by a row of $10^6$ atoms and random configurations of the atoms were calculated based on their respective concentration determined by TRIDYN for each fluence. Subsequently, the total effective number of triplets was determined from the atomic configurations for each fluence. There are six different triplet configurations: horizontal upwards (Pt atom above the two FM atoms), horizontal downwards (Pt atom below the two FM atoms) triplets and four diagonal triplets, see Fig. S2 in SI for a detailed description of the six triplet configurations. The effective number of triplets, which is the difference between the effective number of triplets pointing up or down, gives the strength of the DMI. Evolution of the effective number of triplets with increasing ion-irradiation is shown in Fig. 5. The number of triplets initially increases and reaches a maximum around a fluence of $0.7 - 0.8 \times 10^{16}$ ions/cm², before the number decreases for higher fluences.

*Discussion* - Enhanced X-ray fine-structure absorption spectroscopy has shown that irradiation indeed affects both interatomic distances and atomic arrangement at Co/Pt interfaces [39]. This seems to validate our conclusion that the enhancement of the interfacial DMI is the direct result of intermixing at the interface. However, we expect that the impact of intermixing is rather different at the Ta and the Pt interfaces. Ta in contact with a ferromagnet often quenches the interfacial moments, resulting in a so-called dead layer. For example, Shaw *et al*. measured a dead layer of 0.25 nm for annealed Ta/Co$_{60}$Fe$_{20}$B$_{20}$ [40]. Radiation-induced intermixing at the Ta/CoFeB interface might increase the dead-layer thickness. On the other hand, CoPt and FePt alloys are known to be magnetic [41,42], and the immediate proximity of many ferromagnets with Pt is known to induce a magnetic moment at the interfacial Pt atoms [43]. Recent density functional theory (DFT) calculations, which only considered intermixing of the two atomic layers at the interface, showed that the intermixing at a Co/Pt interface might even increase the magnetic moment of Co [44] as a result of the intermixing at the Co/Pt interface. The same DFT calculations indicate that such intermixing might also reduce the DMI, which requires both spin-orbit coupling and broken inversion symmetry. However, the calculations also suggest that the DMI remains unchanged over a broad range of intermixing after an initial drop. This is caused by the increasing contributions to the DMI from the second Co layer, which is one monolayer away from the interface. The intermixing brings these Co atoms in anisotropic contact with Pt atoms, which provides both the requisite broken inversion symmetry and spin-orbit coupling.

Our Monte-Carlo simulations demonstrate the importance of the contribution of the triplets that are not directly located at the interface, for the DMI strength. We interpret our results as follows: the unirradiated sample already has a certain degree of disorder at the interface, as expected for any such polycrystalline sputtered multilayer, which does not have an atomically flat interface. Thus, the DMI measured prior to the irradiation would correspond to the effective number of triplets for the lowest fluences in our Monte-Carlo simulations in Fig.4. The irradiation then promotes further intermixing of Pt with the magnetic species thereby introducing additional asymmetric exchange interactions at anisotropic triplets, e.g., Co-Pt-Co, that are located further from the nominal interface. This results in an increase of the total effective number of triplets for intermediate fluences and consequentially in the experimentally observed increase of the DMI. The Monte-Carlo simulations show that the number of triplets then decreases for higher fluences, which agrees with the measured drop in the DMI.

Even though the evolution of the total effective number of triplets reflects the general trend of the experimental findings, we do not expect an exact agreement, because additional details beyond the number of triplets need to be considered to reproduce the exact fluence dependence of the DMI. In addition to the fact that the exact initial condition of the CoFeB/Pt interface is not known and the determination of the number of triplets is restricted to a 2D vertical slice through the sample, more subtle details need to be considered, for example: A gradient in the density of triplets and long range effects beyond nearest neighbor interactions can influence the electronic structure.

Finally, this work provides strong evidence that, even though both the DMI and interfacial anisotropy depend on the strength of the spin-orbit coupling, they have a distinctly different dependence on the interface morphology.

*Summary* - He⁺ ion irradiation of Ta/CoFeB/Pt multilayers increases DMI by about 20 % at a fluence of $2 \times 10^{16}$ ions/cm², whereas we find different behavior for the anisotropy for fluences up to $1 \times 10^{16}$ ions/cm² independent of assumptions regarding the saturation magnetization of the CoFeB thickness $H_k^t$ and $H_k^{Ms}$ decrease by about 32 % and 4 % respectively for this fluence. Simulations predict that a large number of atoms from the ferromagnet are displaced within a 0.4 nm thick region from the CoFeB layer and into the Pt layer. We conjecture that the increase in disorder, which includes both roughness and intermixing, results in a concomitant increase in DMI and reduction in anisotropy. Monte-Carlo simulations demonstrate that the general trend of the DMI with fluence can be explained with the total effective number of FM-Pt-FM triplets. As such, we speculate that DMI can be significantly increased by adjustment of deposition conditions that promote an optimum degree of disorder at the interface. Moreover, ion-irradiation opens the possibility to locally tune the DMI down to potentially define regions for the nucleation or annihilation of skyrmions for novel memory devices.

Supplementary Material

Additional information about the anisotropy, Brillouin Light Scattering spectroscopy, TRIDYN and the Monte-Carlo simulations for the FM-Pt-FM triplets is provided.

Acknowledgement

We thank the Ion Beam Center at the Helmholtz-Zentrum Dresden-Rossendorf for the ion irradiation of the samples. We thank Prof. Wolfhard Möller for support on the TRIDYN-simulations. We like to thank one of the referees for comments, which helped us to improve our Monte-Carlo simulations to determine the number of triplets. HTN, TJS and JMS acknowledge support by the DARPA Topological Excitations in Electronics (TEE) program, award No. R18-687-004.

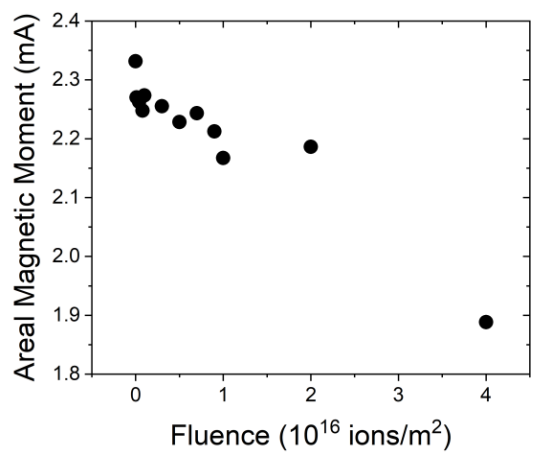

*Figure 1: Fluence dependence of the areal magnetic moment measured by SQUID magnetometry.*

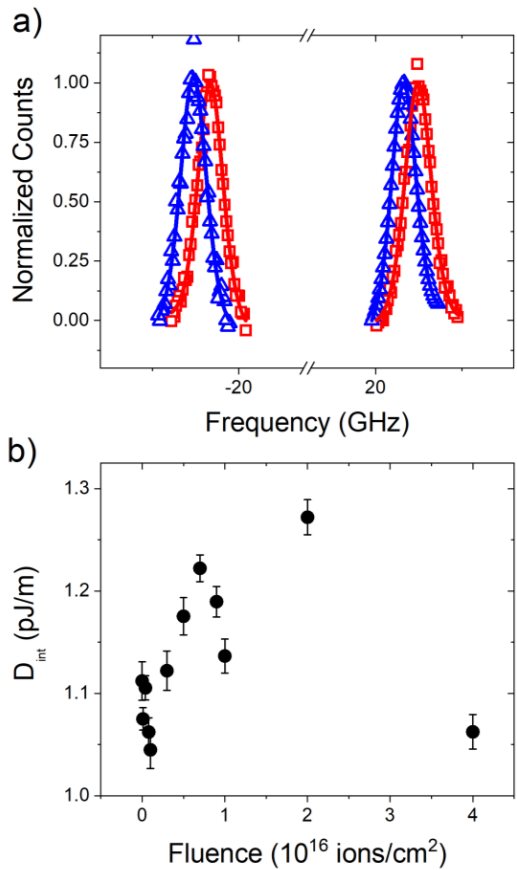

Figure 2: a) Measured BLS for both negative (red squares) and positive field (blue triangles) together with the respective fits (solid lines) for the not irradiated sample. b) Fluence dependence of $D_{int}$, which is calculated from the measured frequency-shift without any assumptions regarding the thickness of the CoFeB layer.

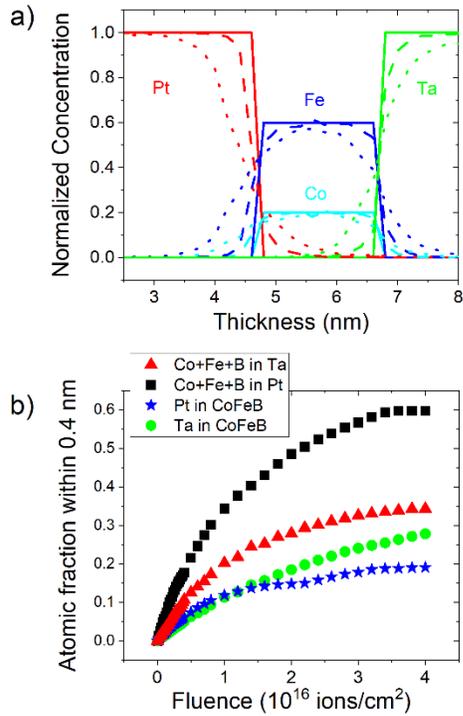

*Figure 3: a) Normalized concentration profile for Ta, Pt, Co and Fe before irradiation (solid line), a fluence of $7 \times 10^{15}$ ions/cm² (dashed line) and $4 \times 10^{16}$ ions/cm² (dotted line). The simulations include both the MgO capping layer and the Si substrate but the above figure focuses only on the layers adjacent to the ferromagnet. b) Fluence dependence of the sum of the atomic fractions of the three element Co, Fe and B in Ta and in Pt within a 0.4 nm thick layer from the respective interface and the percentage of Ta and Pt in a 0.4 nm thick layer at the respective CoFeB interface.*

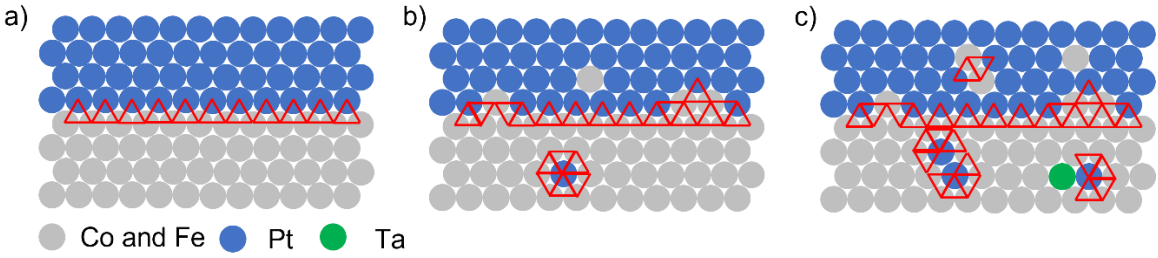

*Fig.4: Illustration of the triplets a) prior to irradiation with an atomically sharp interface, b) with a small fluence and c) with a larger fluence. The tip of the triangles points always to the Pt atom and on side of triangle connect two Fe or Co atoms. Note that triplets with an upwards or downwards orientation contribute to opposite signs of DMI and that the depicted location of the atoms are not the result of the TRIDYN simulations.*

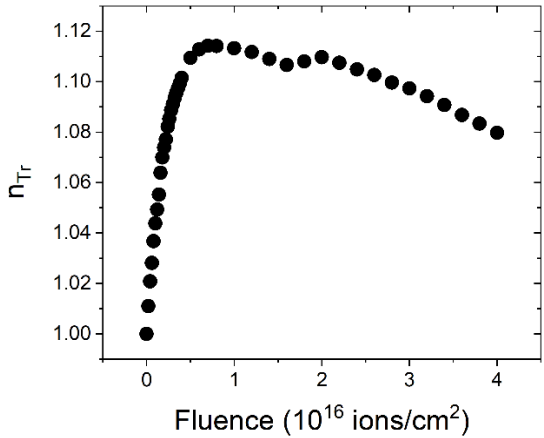

*Figure 5: Evolution of the total normalized number of FM-Pt-FM triplets ($n_{tr}$) with the ion-irradiation fluence. The initial state of the samples prior to irradiation do not have atomically sharp interfaces as in the simulation and as such the initial number of triplets in the experiment is larger than in the simulations.*